\documentstyle[aas2pp4]{article}
\input epsf
\def\plotone#1{\centering \leavevmode
\epsfxsize= 0.8\columnwidth \epsfbox{#1}}

\def\be{\begin{equation}}
\def\ee{\end{equation}}
\def\bea{\begin{eqnarray}}
\def\eea{\end{eqnarray}}

\def\cmm2{{\,\rm cm^{-2}}}
\def\cm2{{\,{\rm cm}^2}}
\def\cmm3{{\,{\rm cm}^{-3}}}
\def\gcmm3{{\,{\rm g\,cm^{-3}}}}

\def\HO{{100h\,{\rm km\,sec^{-1}\,Mpc^{-1}}}}

\def\fun#1#2{\lower3.6pt\vbox{\baselineskip0pt\lineskip.9pt
  \ialign{$\mathsurround=0pt#1\hfil##\hfil$\crcr#2\crcr\sim\crcr}}}

\def\etal{{et al.}}

\def\p3m{P$^3$M}

\def\la{\mathrel{\mathpalette\fun <}}
\def\ga{\mathrel{\mathpalette\fun >}}
\def\fun#1#2{\lower3.6pt\vbox{\baselineskip0pt\lineskip.9pt
  \ialign{$\mathsurround=0pt#1\hfil##\hfil$\crcr#2\crcr\sim\crcr}}}

\font\BF=cmmib10
\def\gam{\hat{\gamma}}
\def\k{{\hbox{\BF k}}}

\def\v{{\hbox{\BF v}}}

\def\gsim{\;\rlap{\lower 2.5pt
 \hbox{$\sim$}}\raise 1.5pt\hbox{$>$}\;}
\def\lsim{\;\rlap{\lower 2.5pt
   \hbox{$\sim$}}\raise 1.5pt\hbox{$<$}\;}

\begin{document}
\twocolumn
\title{Reionization of the Intergalactic Medium
and its Effect on the CMB}

\author{Zolt\'an Haiman}
\affil{NASA/Fermilab Astrophysics Center, Fermi
National Accelerator Laboratory, Batavia, IL 60510}

\author{Lloyd Knox}
\affil{Department of Astronomy and Astrophysics,
University of Chicago, Chicago, IL 60637}

\vspace{.2in}

\begin{abstract}
The bulk of the hydrogen in the universe transformed from
neutral to ionized somewhere in the redshift interval $5 \la z \la
40$, most likely due to ionizing photons produced by an early
generation of stars or mini--quasars.  The resulting free electrons,
interacting with the CMB photons via Thomson-scattering, are a mixed
blessing, providing both a probe of the epoch of the first stars and a
contaminant to the pristine primary anisotropy.  Here we
review our current knowledge of reionization with
emphasis on inhomogeneities and describe the possible connections to
CMB anisotropy.
\end{abstract}


\section{Introduction}

One of the most remarkable observational results in the last three decades in
cosmology is the lack of the Gunn--Peterson trough in the spectra of
high--redshift quasars and galaxies.  This finding implies that the
intergalactic medium (IGM) is highly ionized at least out to $z\sim5$, the
redshift of the most distant known quasars and galaxies\footnote{Currently the
best lower limit on the reionization redshift comes from the detection of
high--$z$ Ly$\alpha$ emission lines, implying $z\geq5.64$ (see discussion
below).}.  Since the primeval plasma recombined at $z\sim1100$, its subsequent
ionization requires some form of energy injection into the IGM, naturally
attributable to astrophysical sources.  The two most popular examples of such
sources are an early generation of stars, residing in sub--galactic size
clusters (hereafter ``mini--galaxies''), or accreting massive ($\sim10^6{\rm
M_\odot}$) black holes in small ($\lsim10^9{\rm M_\odot}$) halos (hereafter
``mini--quasars'').

Reionization is interesting for a variety of reasons.  First, the most
fundamental questions are still unanswered: What type of sources
caused reionization, and around what redshift did it occur?  How were
the reionizing sources distributed relative to the gas? What was the
size, geometry, and topology of the ionized zones, and how did these
properties evolve?  Second, and more directly relevant to the topic of
this review, reionization leaves distinctive signatures on the cosmic
microwave background (CMB) through the interaction between the CMB
photons and free electrons.  Although it may complicate the extraction
of cosmological parameters from CMB data, this ``contamination'' could
yield important information on the ionization history of the IGM.
Through reionization, the CMB is a useful probe of nonlinear processes
in the high--redshift universe.

Thomson scattering from free electrons affects the CMB in several ways.
Because the scattering leads to a blending of photons from initially different
lines of sight, there is a damping of the {\it primary} temperature anisotropy.
On the other hand, a new {\it secondary} anisotropy is generated by what can be
thought of as a Doppler effect: as photons scatter off free electrons, they
pick up some of their peculiar momentum.  Finally, the polarization dependence
of the Thomson cross section creates new polarization from the initially
anisotropic photon field.

The inhomogeneity of reionization affects these processes in two different
ways.  First, the inhomogeneity of the medium affects how the {\it mean}
ionization fraction evolves with time.  This is important since the damping,
Doppler and polarization effects all occur even in the idealized case of
spatially homogeneous reionization.  Secondly, the spatial fluctuations in the
ionization fraction can greatly enhance the contribution from the Doppler
effect at small angular scales.  Polarization and damping are much less
influenced by the inhomogeneity, as we will see below.

The aim of the present review is to summarize our current knowledge of
reionization, describe the possible connections to the CMB, and assess
what we can hope to learn about the high--redshift IGM from
forthcoming observations of the CMB anisotropies.

\section{Reionization}

\subsection{Homogeneous Reionization and the Reionization Redshift}

If the IGM was neutral, its optical depth to Ly$\alpha$ absorption would be
exceedingly high: \\
$\tau_{\rm igm}\sim10^5 (\Omega_{\rm b}h/0.03)
[(1+z)/6]^{3/2}$, wiping out the flux from any source at observed wavelengths
shorter than $\lambda_{\rm Ly\alpha}(1+z)$.  The spectra of high--redshift
quasars and galaxies reveal Ly$\alpha$ absorption by numerous discrete
Ly$\alpha$ forest clouds, separated in redshift, rather than the continuous
Gunn--Peterson (GP) trough expected from a neutral IGM.  The detection of the
continuum flux in--between the Ly$\alpha$ clouds implies that $\tau_{\rm igm}$
is at most of order unity, or, equivalently, an upper limit on the average
neutral fraction of $x=n_{\rm HI}/n_{\rm tot} \lsim10^{-5}$.  The intergalactic
medium is therefore highly ionized.  Note that there is no sharp physical
distinction between the intergalactic medium and the low column density
($N_{\rm HI}\lsim 10^{14}~{\rm cm}^{-2}$) Ly$\alpha$ forest (Reisenegger \&
Miralda-Escud\'e 1995). These Ly$\alpha$ clouds fill most of the volume of the
universe, are highly ionized, and account for most of the baryons (implied by
Big Bang nucleosynthesis) at high redshift (Weinberg et al. 1997; Rauch et
al. 1997).

The energy requirement of reionization can easily be satisfied either by the
first mini--galaxies or mini--quasars\footnote{More exotic possibilities
that we do not discuss here
include primordial black holes (e.g. Gibilisco 1996); cosmic rays (e.g. Nath \&
Bierman 1993); winds from supernovae (e.g. Ostriker \& Cowie 1981); or decaying
neutrinos (Sciama 1993).}. Nuclear burning inside stars releases several MeV
per hydrogen atom, and thin--disk accretion onto a Schwarzschild black hole
releases ten times more energy, while the ionization of a hydrogen atom
requires only 13.6 eV.  It is therefore sufficient to convert a fraction of
$\lsim 10^{-5}$ of the baryonic mass into either stars or black holes in order
to ionize the rest of the baryons in the universe.  In a homogeneous universe,
the ratio of recombination time to Hubble time in the IGM is $t_{\rm
rec}/t_{\rm Hub}\sim [(1+z)/11]^{-3/2}$ so that each H atom would only
recombine a few times at most until the present time. Unless reionization
occurred at $z\gg 10$, recombinations therefore would not change the above
conclusion, and the number of ionizing photons required per atom would be of
order unity.  Based on three-dimensional simulations, Gnedin \& Ostriker (1997)
argued that the clumpiness of the gas increases the average global
recombination rate by a factor $C\equiv<\rho^2>/<\rho>^2\sim 30-40$ (the
averages are taken over the simulation box).  This would imply a corresponding
increase in the necessary number of ionizing photons per atom.  However, in a
more detailed picture of reionization in an inhomogeneous universe (Gnedin
1998, Miralda-Escud\'e et al. 1999), the high-density regions are ionized at a
significantly later time than the low density regions.  If this picture is
correct, then one ionizing photon per atom would suffice to reionize most of
the volume (filled by the low-density gas); recombinations from the
high-density regions would only contribute to the average recombination rate at
a later redshift.

\begin{figure}[bthp]
\plotone{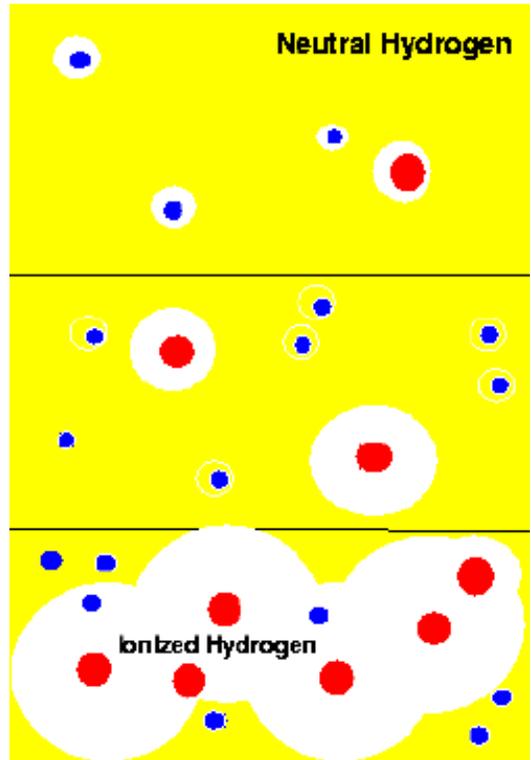}
\caption[]{\baselineskip=10pt
Schematic stages of reionization by early mini--galaxies. The small
dots denote small halos with $T_{\rm vir}<10^4$K,  the
larger dots correspond to larger halos with $T_{\rm vir}>10^4$K.
Top panel -- $z\sim40$: The first halos form, both small and large halos
create HII zones.  Middle panel -- $z\sim 20$: The UV background suppresses
star-formation in small halos, only large halos can produce HII zones.  Bottom
panel -- $z\sim 15$: the ionized zones from the large halos overlap.}
\label{fig:bubbles} 
\end{figure}

The reionization redshift can be estimated from first principles.  In the
simplest picture [first discussed by Arons \& Wingert (1972) in the context of
quasars], a homogeneous neutral IGM is percolated by the individual HII regions
growing around each isolated UV source. The universe is reionized when the
ionized bubbles overlap, i.e. when the filling factor of HII regions reaches
unity\footnote{The expansion of spherical ionization fronts around steady
sources can be calculated analytically (Shapiro \& Giroux 1987).}.  Several
authors studied this problem (Carr et al. 1984, Couchman \& Rees 1986,
Fukugita \& Kawasaki 1994, Meiksin \& Madau 1993; Shapiro et al. 1994,
Tegmark et al. 1994; Aghanim et al. 1996; Haiman \& Loeb 1997).  Specifically,
Haiman \& Loeb (1998a) have used the Press--Schechter formalism to describe the
formation of halos that potentially host ionizing sources, in order to estimate
the redshift at which overlap occurs.  In this study, star and quasar black
hole formation was allowed only inside halos that can cool efficiently via
atomic hydrogen, since molecular hydrogen would be photodissociated earlier, as
argued by Haiman, Abel \& Rees (1999) and Haiman, Rees \& Loeb (1997).  This
requirement translates to a minimum virial temperature of $\sim10^4$K, or halo
masses above $\sim10^{8}{\rm M_\odot}[(1+z)/11]^{-3/2}$ (see
Fig.~\ref{fig:bubbles}).  For both types of sources, the total amount of light
they produce was calibrated using data from redshifts $z\la 5$. The efficiency
of early star formation can be estimated from the observed metallicity of the
intergalactic medium (Songaila \& Cowie 1996, Tytler~et~al.~1995), utilizing
the fact that these metals and the reionizing photons likely originate from the
same stellar population.  Similarly, the efficiency of black hole formation
inside early mini--quasars can be constrained to match the subsequent evolution
of the quasar luminosity function at redshifts $z\la 5$ (Pei 1995).  These
assumptions generically lead to a reionization redshift $8\lsim z \lsim
15$. The uncertainty in the redshift reflects a range of cosmological
parameters ($H_0, \Omega_m, \Omega_b$), CDM power spectra ($n$, $\sigma_8$),
and efficiency parameters for the production and escape fraction of ionizing
photons.  In the stellar reionization scenario, the predicted redshift could be
above or below this range, if the initial mass function was strongly biased
(relative to Scalo 1986) towards massive or low-mass stars.

\begin{figure}[bthp]
\plotone{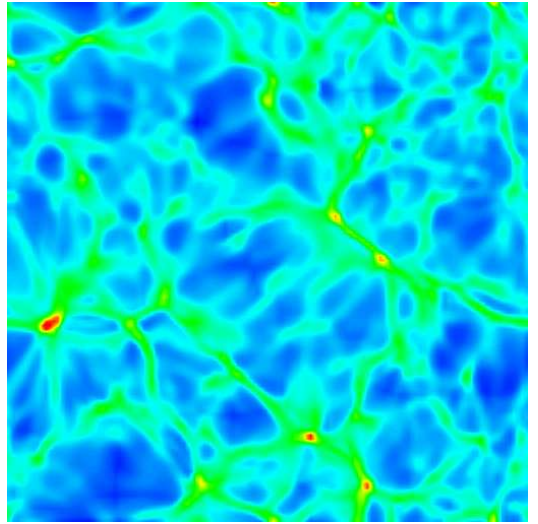}
\caption[]{A two--dimensional slice of log($\delta_{\rm b}$), the 
baryonic overdensity at redshift $z$=5 in a three--dimensional simulation 
(see Zhang et al. 1998).}
\label{fig:3d} 
\end{figure}

\subsection{Importance and Treatments of Inhomogeneities}

Although the naive picture of overlapping HII spheres may give a good estimate
for the ``overlap redshift'', it is likely to be a crude and incomplete
description of the real process of reionization.  The ionizing sources are
expected to be located in the highly overdense regions that form the complex
large scale structure of the universe - along filaments or sheets, or at their
intersections (see Fig.~\ref{fig:3d}).  The ionizing radiation will have to
cross the local dense structures before propagating into the IGM, which itself
has significant density fluctuations.  Several important consequences of these
inhomogeneities can be noted.  First, reionization could occur ``outside-in''
(Miralda-Escud\'e et al. 1999), if the radiation from a typical source escapes
through a relatively narrow solid angle from its host, without causing
significant local ionization. In this case, the low--density regions could be
ionized significantly earlier than the high--density regions.  An important
effect in this case could be the shadowing of ionizing radiation by the dense
concentrations within the IGM.  The same type of shadowing effect was found to
be important at low redshift.  Based on the number of Ly$\alpha$ absorbers,
Madau, Haardt \& Rees (1998) have concluded that as much as $\sim 50$\% of the
UV flux is absorbed by these systems.  The redshift at which the IGM becomes
optically thin to the ionizing continuum is therefore significantly delayed
relative to the overlap epoch, when the ionized zones first percolate.  An
alternative possibility is that the ionizing sources are densely spaced along
filaments, and their radiation ionizes most of the dense filament before
escaping into the IGM.  In this ``inside-out'' case, the description of ionized
zones surrounding their host source would be more applicable.  An important
feedback during and after reionization is that the newly established UV
background can photo--evaporate the gas from small ($v_{\rm circ}\lsim10$)
halos (Shapiro, Raga \& Mellema 1997; Barkana \& Loeb 1999). A related point,
important for the detection of the GP trough, is that at the time of overlap of
the HII zones, each individual HII region could still have a non--negligible
Ly$\alpha$ optical depth, due to its residual HI fraction.  The GP trough
therefore disappears only from the spectra of individual sources 
located below
the later redshift when the average Ly$\alpha$ optical depth of the IGM drops
below unity (Haiman \& Loeb 1998b; Miralda-Escud\'e et al. 1999; Shapiro et
al. 1987).  Finally, the inhomogeneity of the reionized IGM leads to generation
of anisotropies in the CMB which are strongly suppressed in the homogeneous
case, as we shall see below.

Reionization is now also being addressed by three-dimensional simulations,
which can eventually shed light on some of the above issues.  The first such
studies approximated radiative transfer by assuming an isotropic radiation
field, but were able to take into account the inhomogeneities in the gas
distribution (Ostriker \& Gnedin 1996; Gnedin \& Ostriker 1997).  The resulting
reionization redshift of $z\sim7$ in a $\Lambda$CDM model is in good agreement with the
semi--analytical estimates, given the differences in the assumed cosmology,
power spectra, and star formation efficiency.  However, an isotropic radiation
field is necessarily a crude approximation, at least in the beginning stages of
reionization, when each source still has its own isolated ionized zone.  In
order to understand the progressive overlap of ionized zones, including effects
such as self-shielding and shadowing by dense clumps, it is necessary to
incorporate radiative transfer into three dimensional simulations (Norman et
al. 1998). The numerical algorithms for the first such attempts are described
in Abel, Norman \& Madau (1998) and Razoumov \& Scott (1998).  Results from an
approximate treatment (Gnedin 1998) also support the outside--in picture of
reionization described above.

\subsection{The Nature of Reionizing Sources}

Several authors have emphasized that the known population of quasars or
galaxies provide $\sim$10 times fewer ionizing photons than necessary to
reionize the universe (see, e.g. Shapiro et al. 1994, or Madau et al. 1999, and
references therein).  The two most natural candidates for undetected ionizing
sources are the mini--galaxies and mini--quasars expected to be associated with
small ($\sim10^{8-10}~{\rm M_\odot}$) dark halos at $z\sim10$.  As there are no
compelling {\it ab initio} theoretical arguments to strongly favor one type of
source over the other, the best hope is to distinguish these two possibilities
from observations.  One expects at least three major differences between
mini--galaxies and mini--quasars: their spectra, their absolute brightness (or,
alternatively, space density), and the angular size of their luminous regions.

The flux from stellar populations drops rapidly with frequency above the
ionization threshold of hydrogen, while the spectra of mini--quasars are
expected to be harder and extend into the X--ray regime.  Stars could therefore
not reionize HeII, while quasars with typical spectra (Elvis et al. 1994) would
reionize HeII at approximately the same redshift as H (Haiman \& Loeb 1998a).
However, recent high-resolution spectra of $z\sim3$ quasars have shown a widely
fluctuating HeII/HI optical depth (Reimers et al. 1997).  The authors
interpreted this observation as evidence for HeIII regions embedded in an
otherwise HeII medium, i.e. a detection of the HeII reionization epoch (see
also Wadsley et al. 1998).  If this interpretation is verified by future data,
it would constrain the number of mini--quasars with hard spectra extending to
X--rays at high redshifts (Haiman \& Loeb 1998b; Miralda-Escud\'e 1998;
Miralda-Escud\'e \& Rees 1994).  At present, a plausible alternative
interpretation of the HeII observations is that the observed HeII optical depth
fluctuations are caused by statistical fluctuations in the IGM density or the
ionizing background flux (Miralda-Escud\'e et al. 1999), rather than by the
patchy structure of HeII/HeIII zones.

The X--ray background (XRB; Miyaji et al. 1998; Fabian \& Barcons 1992) might
provide another useful constraint on mini--quasar models with hard spectra
(Haiman \& Loeb 1998c).  These models overpredict the {\it unresolved
flux} by a factor of $\sim 2-7$ in the 0.1--1 keV range.  If
an even larger fraction of the XRB will be resolved into low--redshift AGNs in
the future, then the XRB could be used to place more stringent constraints on
the X-ray spectrum or the abundance of the mini--quasars.

A second distinction between galaxies and quasars can be made based on their
absolute luminosity and space density. On average, based on their $z<5$
counterparts, high--redshift mini--quasars are expected to be roughly $\sim
100$ times brighter, but $\sim 100$ times more rare, than galaxies. At the 
overlap
epoch, HII regions from mini--quasars would therefore be larger but fewer than
those from mini--galaxies.  The typical size and abundance of HII regions has
at least
three important implications.  First, it affects the CMB anisotropies;
the larger the HII regions, the larger the effect on the CMB (Gruzinov
and Hu 1998, Knox et al. 1998), as we
will see below.  Second, large enough HII
regions ($\gsim 1$Mpc) would allow gaps to open up in the GP trough; the flux
transmitted through such gaps around very bright quasars could be detectable
(Miralda-Escud\'e et al. 1999).  Third, if sources form inside the rare,
high-$\sigma$ peaks, they would be strongly clustered (Knox et al. 1998).  
This clustering would
increase the effective luminosity of each source, and would enhance both of the
effects above.

Finally, a constraint on the number of mini--quasars at $z>3.5$ can be derived
from the Hubble Deep Field (HDF).  Mini--quasars are expected to appear as
faint point--sources in the HDF, unless their underlying extended host galaxies
are resolved. The properties of faint {\it extended} sources found in the
Hubble Deep Field (HDF) agree with detailed semi--analytic models of galaxy
formation (Baugh et al. 1998).  On the other hand, the HDF has revealed only a
handful of faint {\it unresolved} sources, and none with the colors expected
for high redshift quasars (Conti et al. 1999).  The simplest mini--quasar
models predict the existence of $\sim10-15$ B--band ``dropouts'' in the HDF,
inconsistent with the lack of detection of such dropouts up to the $\sim50\%$
completeness limit at $V\approx 29$.  To reconcile the models with the data, a
mechanism is needed that suppresses the formation of quasars in halos with
circular velocities $v_{\rm circ} \lsim 50-75~{\rm km~s^{-1}}$ (Haiman, Madau
\& Loeb 1999).  This suppression naturally arises due to the photo-ionization
heating of the intergalactic gas by the UV background after reionization.
Forthcoming data on point--sources from NICMOS observations of the HDF (see
Thompson et al. 1998) could further improve these constraints.

\subsection{When was the Universe Reionized?}

As described above, theoretical expectations place the reionization redshift at
$8\lsim z\lsim 15$.  What about observations?  At present, the best {\em lower}
limit in the reionization redshift comes from the detection of high--redshift
Ly$\alpha$ emitters (Hu et al. 1998; Weymann et al. 1998).  As argued by
Miralda-Escud\'e (1998), the damping wing of Ly$\alpha$ absorption by a neutral
IGM has a large residual optical depth that would severely damp any Ly$\alpha$
emission line.  Currently the highest redshift at which a Ly$\alpha$ emitter is
seen is $z=5.64$; the existence of this object implies that reionization
occurred prior to this redshift (see Haiman \& Spaans 1998). Note that HII
regions around individual sources, excepting only the brightest quasars, would
be too small to allow the escape of Ly$\alpha$ photons, because the damping
wings of the GP troughs around the HII region would still overlap.

The best {\em upper} limit on the reionization redshift can be obtained from
the CMB anisotropy data. Several experiments have revealed a rise in the power
on small angular scales, and a drop at even smaller scales (e.g.,
Bond \etal~1999), showing evidence
for the first Doppler peak expected from acoustic oscillations in the
baryon-photon fluid prior to
recombination (e.g., Hu \& Sugiyama 1994).
These observations can also be used to
set limits on the electron scattering optical depth (or the corresponding
reionization redshift) that would suppress the Doppler peak.  From a
compilation of all the existing measurements, Griffiths et al. (1998) have
derived the stringent (although model-dependent)
constraint $z\lsim 40$ on the reionization redshift -- if
reionization occurred earlier, the electron scattering optical depth would have
reduced the amplitude of the anisotropies below the observed level.  Note that
another upper limit results from the spectral distortion of the CMB 
caused by
scattering on the reionized IGM, which is constrained by the upper limit on the
Compton $y$--parameter measured by COBE. The inferred upper limit on the
reionization redshift, however, is much weaker, $z\lsim 400$ for typical
parameters in a low--density universe (Griffiths et al. 1998, see also Stebbins
\& Silk 1986).  In summary, present observations have narrowed down the
possible redshift of the reionization epoch to $6\lsim z\lsim 40$, a relatively
narrow range that is in good agreement with the theoretical predictions
described above.

\subsection{Future Observational Signatures}

Further observational progress in probing the reionization epoch could come
either from the {\it Next Generation Space Telescope} ({\it NGST}), or from
more precise measurements of the CMB anisotropies by 
{\it MAP}\footnote{http://map.gsfc.nasa.gov} and 
{\it Planck}\footnote{http://astro.estec.esa.nl/SA-general/Projects/Planck}.  
{\it NGST}, scheduled for launch in 2007, is expected to reach the
$\sim1$nJy sensitivity\footnote{See the {\it NGST} Exposure Time Calculator at
http://augusta.stsci.edu.} required to detect individual sources to $z\sim10$,
and to perform medium--resolution spectroscopy to $z\sim8$.  If reionization
occurred close to the low end of the allowed redshift range, $6\lsim z\lsim8$,
then it may be possible to infer the redshift directly from the spectra of
bright sources.  One specific method relies on the spectrum of a bright source
just beyond the reionization redshift, so that the individual Ly$\alpha$,
Ly$\beta$, and other GP troughs do not overlap in frequency, leaving gaps of
transmitted flux (Haiman \& Loeb 1998b).  The measurement of this transmitted
flux would be possible with {\it NGST}, despite absorption by the
high--redshift Ly$\alpha$ forest, as long as the number density of absorbers
does not rise much more steeply with redshift than an extrapolation of the
current $z<5$ data would imply (Fardal et al. 1998). If reionization is
gradual, rather than abrupt, than this method would measure the redshift at
which the GP optical depth drops to near unity, i.e. the final stages of the
reionization epoch.  An alternative signature to look for would be the precise
shape of the damping wing of Ly$\alpha$ absorption from the neutral IGM along
the line of sight to the source (Miralda-Escud\'e 1998).  The shape, if
measured by high--resolution spectroscopy, could be used to determine the total
optical depth of the IGM. The caveat of this method is the inability to
distinguish the neutral IGM from a damped Ly$\alpha$ absorber along the line of
sight, close to the source.

An alternative signature is the background Ly$\alpha$ emission from the
reionization epoch.  Recombinations are slow both at high redshifts, when the
IGM is still neutral, and at low redshifts when the IGM density is low.  As
shown by simulations (Gnedin \& Ostriker 1997), the global recombination rate
has a pronounced peak around the reionization epoch.  The resulting recombinant
Ly$\alpha$ emission has been computed by Baltz, Gnedin \& Silk (1998), and a
detailed detectability study has shown that this signal could be measured by
{\it NGST}, or perhaps even by {\it HST} (Shaver et al. 1999).

Yet another signature could result from the 21 cm hyperfine transitions in the
IGM before reionization. Prior to reionization, the excitation of the 21 cm
line in neutral HI atoms depends on the spin temperature.  The coupling of the
spin temperature to that of the CMB is determined by the local gas density,
temperature, and the radiation background from the first mini--galaxies and
mini--quasars.  In general, the 21 cm line could be seen either in absorption
or emission against the CMB, and could serve as a 'tomographic' tool to
diagnose the density and temperature of the high redshift neutral gas (see,
e.g. Madau et al. 1997; Scott \& Rees 1990).  If reionization occurred at
$6\lsim z \lsim 10$, the redshifted 21 cm signals would be detectable by the
Giant Metrewave Radio Telescope; a study of the effect at higher redshifts
would be possible with next generation instruments such as the THousand Element
Array, or the Square Kilometer Array (Shaver et al. 1999).

The reionization redshift could turn out to be closer to the high end of the
allowed range, $10\lsim z\lsim40$, rendering direct detection of emission from
the reionization epoch implausible (except perhaps the 21cm signal).
There is, however, a fortunate ``complementarity'', since the electron
scattering optical depth increases with reionization redshift.  Due
to the effect on the polarization power spectrum on
large angular scales, {\it MAP} and
{\it Planck} may be able to discern an electron 
scattering optical depth as small as a few
percent (Zaldarriaga et al. 1997, Eisenstein et al. 1998, 
Prunet et al. 1998).  By temperature anisotropy data alone,
they may be able to determine an optical depth as small as $\sim 20\%$
due to the damping effect mentioned earlier---and to be discussed in
more detail later.

The reionizing sources may also change the spectral shape of the CMB.  The dust
that is inevitably produced by the first type II supernovae, absorbs the UV
emission from early stars and mini--quasars and re-emits this energy at longer
wavelengths.  Loeb \& Haiman (1997) have quantified the resulting spectral
distortion in Press--Schechter type models, assuming that each type II
supernova in a Scalo IMF yields ${\rm 0.3M_\odot}$ of dust with the
wavelength-dependent opacity of Galactic dust, uniformly distributed throughout
the intergalactic medium.  Under these assumptions, the dust remains cold
(close to the CMB temperature), and its emission peaks near the CMB peak. The
resulting spectral distortion can be expressed as a Compton $y$--parameter
$\sim 10^{-5}$, near the upper limit derived from measurement of the CMB
spectrum by COBE (Fixsen et al. 1996).  A substantial fraction
($\sim10$--$50$\%) of this total $y$--parameter results simply from the direct
far-infrared emission by early mini--quasars and could be present even in the
absence of any intergalactic dust.

Inhomogeneities in the dust distribution could change these conclusions.
Instead of being homogeneously mixed into the IGM, the high--redshift dust may
remain concentrated inside or around the galaxies where it is produced.  In
this case, the average dust particle would see a higher flux than assumed in
Loeb \& Haiman (1997), so that the dust temperature would be higher, and dust
emission would peak at a shorter wavelength.  The magnitude of the spectral
distortion would be enhanced, and may account for the recently discovered
cosmic infrared background (Puget et al. 1996, Schlegel et al. 1998, Fixsen
et al. 1998, Hauser et al. 1998) as shown by the semi--analytic
models of Baugh et al. (1998).  An angular fluctuation in the magnitude of the
distortion would also be expected in this case, reflecting the discrete nature
of the sources contributing to the effect.

As seen above, reionization has several consequences; in the rest of this
review we focus on the connection to the CMB, and the effects of
inhomogeneities.

\section{Effect on the CMB}

In this section, we give an in-depth treatment of the effects
on the CMB photons of Thomson-scattering off of the free electrons
produced by reionization.  After defining the optical depth and
giving its dependence on the redshift of reionization and
cosmological parameters, we discuss the three separate effects:
damping, Doppler and polarization generation.  Then we derive
the evolution equations from the Boltzmann equation, argue
that the Doppler effect is the most important {\it inhomogeneous}
effect, and calculate the resulting power spectrum for several
simple models of inhomogeneous reionization (IHR).

The probability of a photon scattering in the time interval
from some initial time $t_i$ to the present, $t_0$ 
is given by $1-e^{-\tau(t_0)}$ where 
\be
\tau(t_0) \equiv \int_{t_i}^{t_0} \sigma_T n_e(t) dt
\ee
is the optical depth to Thomson scattering and $\sigma_T$ is the
Thomson cross-section.
If one assumes a step-function transition from a neutral to an
ionized IGM at a redshift of $z_{\rm ion}$, the mean optical depth is
given by (e.g., Griffiths et al., 1998)
\be
\label{eqn:depthofzion}
\tau(z_{\rm ion}) = \tau^*/\Omega_0\left[\left(1-\Omega_0 + \Omega_0
\left(1+z_{\rm ion}\right)^3\right)^{1/2}-1\right]
\ee
where
\be
\tau^* = {H_0 \Omega_b \sigma_T \over 4 \pi G m_p} \times \left(1-Y/2\right)
\simeq 0.033 \Omega_b h,
\ee
$\Omega_b$ and $\Omega_0$ are the density of baryonic matter and of all matter
respectively, in units of the critical density today, $H_0$ is the Hubble constant,\\
$H_0 = \HO$, $G$ is Newton's  constant,
$m_p$ is the proton mass and $Y$ is the mass fraction of baryons in helium.  The
final equality assumes $Y=0.24$.
To get equation~\ref{eqn:depthofzion}, 
we assumed that $n_e\propto(1+z)^3$, and adopted
the line element, $dz/dt$, of a
spatially flat universe with only non-relativistic matter
and a cosmological constant.

Note that for $z_{\rm ion}=5.64$ (the observational lower bound),
$\Omega_b h^2 = 0.02$ and $h=0.65$, we get
$\tau = 0.016$ for $\Omega_0 = 1$ and $\tau = 0.049$ for
$\Omega_0 = 0.3$.
For $z_{\rm ion}$ only slightly larger than this limiting value,
it is a good approximation to rewrite equation~\ref{eqn:depthofzion}
as
\be
\tau(z_{\rm ion}) = 0.037/\sqrt{\Omega_0}
\left({1+z_{\rm ion} \over 11}\right)^{3/2}
\left({\Omega_b h^2 \over 0.02}\right) \left({0.65 \over h}\right).
\ee
Thus we see that the fraction of CMB photons that have
been scattered may be very small, but that this fraction
grows fairly rapidly with increasing redshift.
The increase in optical depth for fixed $z_{\rm ion}$
with increasing $1-\Omega_0$ is due to the fact that the proper
time from the present to $z_{\rm ion}$ increases as
the cosmological constant increases.

\subsection{Simple Arguments}

Here we give simple arguments about the nature of the
damping, Doppler and polarization effects.  Before doing so, it is
useful to define the two-point correlation function, $C(\theta)$ 
which is perhaps the most important statistical property of the CMB
fluctuations:
\be
C(\theta) \equiv \langle \Delta (\hat \gamma_1,{\bf x},\eta_0)
\Delta (\hat \gamma_2,{\bf x},\eta_0)  \rangle ; \cos \theta \equiv
\hat \gamma_1 \cdot \hat \gamma_2
\ee
where $\langle ... \rangle$ denotes ensemble average.
The same information is contained in its 
Legendre transform, the angular power spectrum,
$C_\ell$:
\be
C(\theta)=\sum_\ell {2\ell+1\over 4\pi} C_\ell P_\ell(\cos\theta).
\ee
Roughly speaking, $C_l$ is the power in modes with wavelength
$\pi/l$ so that $l = 180$ corresponds to about a degree.
If the fluctuations are Gaussian-distributed and statistically
isotropic, all other statistics can be derived from $C_l$.

\subsubsection{Damping}

For an initially homogeneous and isotropic photon field, there
is just as much probability to scatter into a line of sight
as there is to scatter out of it.  Therefore, the optical depth
to scattering produces no net effect.  If, however, there is initial
anisotropy, then this anisotropy is damped.   Consider a line
of sight along which the temperature differs by $\Delta T$ from
the mean $\bar T$ in the absence of damping.  If damping is
present, then the temperature is changed to
\bea
\label{eqn:simpledamping}
\bar T +\Delta T &\rightarrow& \left(\bar T + \Delta T\right) -
\left(\bar T+\Delta T\right)\left(1-e^{-\tau}\right) +
\bar T\left(1-e^{-\tau}\right) \nonumber \\
&\rightarrow& \bar T + \Delta T e^{-\tau}
\eea
This equation expresses the fact that the final temperature is given
by the initial temperature, reduced by the photons that have
been lost, and increased by the photons that have been scattered
in from other lines of sight.  Since the photons that have been
scattered in come from many different lines of sight, their
average temperature is very nearly the global average---an
assumption made in equation~\ref{eqn:simpledamping}.
The net result is that $\Delta T \rightarrow \Delta T e^{-\tau}$ and therefore
$C(\theta) \rightarrow C(\theta)e^{-2\tau}$ or $C_l \rightarrow C_le^{-2\tau}$.
This simple calculation would imply that the damping is
independent of scale, which is not exactly true.
Our assumption that the average $T$ along lines of sight that
scattered in equals the global average holds only when the distance
{\it between} the scattering surfaces is much larger than the
length scale of interest.  At very large angular scales, the
damping does not occur, as one would expect from simple causality
considerations.

More precisely, the damping factor is $l$-dependent and, defined
via $C_l = R_l^2 C_l^{\rm primary}$, is given 
by the fitting formula of Hu \& White 1997:
\be
R_l^2 = {1-\exp\left(-2\tau\right) \over a+c_1 x +c_2 x^2 + c_3 x^3
+ c_4 x^4} + \exp\left(-2\tau\right),
\ee
with $x=l/(l_r+1)$ and $c_1 = -0.267, c_2 =0.581, c_3 = -0.172$ and
$c_4 = -0.0312$.
The characteristic angular scale $l_r$ is roughly the angular
scale subtended by the horizon at the new last-scattering
surface and is given approximately by (Griffiths et al. 1998, Hu \&
White 1997):
\be
l_r = \left(1+z_{\rm ion}\right)^{1/2}\left(1+0.84\ln \Omega_0\right)-1.
\ee
If not for the $l$-dependence, the damping effect would be completely
degenerate with the amplitude of the primary power spectrum; i.e.,
their effects would be indistinguishable.  The similar response
to amplitude and $\tau$ over a large range of $l$ makes them
approximately degenerate and is the reason why the optical
depth can only be determined to about 10\% (Zaldarriaga et al. 1997)
or possibly worse (Eisenstein et al. 1998) based on
temperature anisotropy alone.

\subsubsection{Polarization}

The Thomson scattering {\it differential}
cross-section is polarization-dependent, and therefore the
scattered radiation may be polarized even if the incident
radiation is not.  However, by symmetry considerations alone
one can see that initially isotropic radiation will not
become polarized\footnote{Unless the electron spins are aligned,
perhaps by a magnetic field.}.
In fact, it is easy to show that the particular
anisotropy required for the creation of polarization is
a quadrupole moment.  This is essentially due to the $\cos 2\theta$
dependence of the differential cross-section.

The Fourier analogue of the angle-distance relation can be written as
$\ell \sim (\eta - \eta_*) k$ where $k$ is the comoving wave-number
that projects from conformal time $\eta_*$ into multipole moment $\ell$ at
conformal time $\eta$.  Conformal time is related to proper time
by $d\eta = dt/a$ where $a$ is the scale factor of the expansion.
Applying this relationship twice, and using the
fact that polarization is generated by the quadrupole moment,
$\ell = 2$, we find
\be
\ell_p \sim (\eta_0 - \eta_{\rm ion})\left({2 \over \eta_{\rm ion}-
\eta_{\rm LSS}}\right) \simeq 2 \left(\sqrt{z_{\rm ion} +1} -1\right).
\ee
For the last equality, we assumed a matter-dominated Universe and that
$\eta_{\rm ion} >> \eta_{\rm LSS}$.  Thus we expect a peak in the
polarization power spectrum near $\ell_p \sim 2z_{\rm
  ion}^{1/2}$.

Determining the location of this peak allows one to determine the
epoch of reionization.  The amplitude of the peak is proportional to
the optical depth, and thus also helps in determining $z_{\rm ion}$.
Unfortunately, the signal is very weak and present only at 
large angular scales,
e.g., $\ell_p \simeq 5$ for $z_{\rm ion} =10$.  
Since there are only $2l+1$ independent modes from which to
determine $C_l$, sample variance is worst at low $l$.  Also,
the large-scale features of polarization maps reconstructed
from time-ordered data are likely to be those most sensitive
to systematic errors.  Finally, polarized galactic foregrounds
(dust and synchrotron) are expected to be most troublesome
at small $l$ (Bouchet et al. 1998, Knox 1998). 
However, a benefit of the low-$\ell$ location of this signature is
that if it can be measured, its interpretation will not be complicated
by the inhomogeneity of the reionization, which is a much
smaller-scale phenomenon.

If the systematic errors are negligible, the reionization feature
in the polarization power spectrum can be used to discern
very small optical depths.  Forecasts of parameter determination
by Eisenstein et al. (1998) using both the temperature anisotropy 
and polarization data are for one sigma errors on $\tau$ of 
0.022 for {\it MAP} and  $0.004$ for {\it Planck}.  Also
see Zaldarriaga et al. (1997) for similar results.  Note that
these sensitivities, especially in the case of {\it Planck}, are
high enough to detect the effects of reionization {\it at any
epoch} given that we know it happened prior to $z=5.64$.

\subsubsection{Doppler Effect}

The contribution to $\Delta T/T$ at location ${\bf x}_0$, in direction
$\hat \gamma$, at conformal time $\eta_0$ (today) is given
by
\be
\label{eqn:Doppler}
{\Delta T \over T}\left({\bf x}_0,\hat \gamma,\eta_0\right)
= \sigma_T\int_{\eta_{\rm ion}}^{\eta_0} d\eta
n_e\left({\bf x}\right)\hat\gamma \cdot \v_e({\bf x}) a 
\ee
where 
${\bf x} = {\bf x}_0 + \hat \gamma\left(\eta_0 - \eta\right)$.
As one might expect, it is proportional to the line-of-sight integral
of the parallel component of electron velocity, ${\bf v}_e$ times
the number density of scatterers, $n_e$.

As Sunyaev (1978) and Kaiser (1984) pointed out,
in the homogeneous case this contribution to $\Delta T \over T$ is
suppressed by cancellations due to oscillations in ${\bf v}_e$.  One
way to think of these cancellations is that
photons get nearly opposite Doppler shifts on different sides of a
density peak, which is a consequence of potential flows generated by
gravitational instability.  
The cancellations are less complete
at large angular scales ($l \la 100$) and there is a measurable
effect which is exactly taken into
account in Boltzmann codes such as CMBFAST\footnote{
http://www.sns.ias.edu/~matiasz/CMBFAST/cmbfast.html}.   If speed is
critical, the fitting formulae of Griffiths et al. (1998)
can be employed for an approximate treatment.

The cancellations at small scales can be greatly reduced by the modulation
of the number density of free electrons which we can write as $n_e =
x_e n_p$ where  
$x_e$ is the ionization fraction and $n_p$ is the number density of 
all protons, including those in H and He.
This modulation occurs even in the case of
homogeneous reionization (spatially constant $x_e$) due to spatial
variations in $n_p$.  Because $v_e$ is zero to zeroth order, and $n_p$
is uniform to zeroth order, this is a second-order effect.  That this
second order effect could dominate the first-order one (which is small
due to cancellations) was pointed out by Ostriker and Vishniac and is
known as the Ostriker-Vishniac effect (Ostriker \& Vishniac 1986,
Vishniac 1987).  Subsequent systematic study of all 2nd order contributions
have shown that it is the dominant second-order 
contribution as well (Hu et al. 1994, Dodelson \& Jubas 1995). 

Another source of modulation of the number density of free electrons
is spatial variations in $x_e$, i.e., IHR.  As a photon streams
towards us from the last-scattering surface, it may pick up
a Doppler ``up kick'' on one side of an overdensity, but avoid the
canceling ``down kick'' on the other because the IGM there is still neutral.  
How effectively the cancellation is avoided depends on the matching
between the typical sizes of the ionized domains and the correlation
length of the velocity field, as we will discuss below.

\subsection{Evolution Equations}

We now derive equation~\ref{eqn:Doppler}, as well as a more general
expression which includes the contribution from damping as well.
More complete treatments of the evolution of a photon distribution
function in an expanding, inhomogeneous Universe can be found
in, e.g., Ma \& Bertschinger (1995).

The Boltzmann equation governing the evolution of the photon
phase-space distribution function, $f$, is simply ${d f \over dt} =
C$, where $C$ takes into account the effect of collisions---in this
case with electrons via Thomson scattering.  The Thomson scattering
cross-section is independent of frequency and therefore does not
induce spectral distortions; the effect on the
distribution function can be fully described by the change in
brightness.  Thus we use the brightness perturbation variable defined
by $\Delta \equiv (\partial f_0 /\partial q)^{-1}f_1$ where $q$ is the
comoving photon momentum, and the phase-space distribution function
has been partitioned into a homogeneous and inhomogeneous part,
$f=f_0+f_1$.  For a Planckian $f$ with thermal fluctuations,
$\Delta T$, about a mean temperature of $T$, $\Delta = 4 \Delta T/T$.

Expanding the total derivative on $f$, keeping terms to first order
in the perturbation variables and Fourier transforming the result, the
Boltzmann equation becomes:
\bea
\label{eqn:Boltz}
\dot {\tilde \Delta}(\k,\gam,\eta) + ik\mu {\tilde \Delta}(\k,\gam,\eta)+ 
{2\over 3}\dot h(\k,\eta) & & \nonumber\\
+{2\over 3} \left(3\dot h_{33}(\k,\eta)-\dot h(\k,\eta)\right)P_2(\mu) &=& \nonumber \\
\int d^3x e^{i{\bf k}\cdot{\bf x}}
\dot \tau ({\bf x},\eta)S({\bf x},\hat \gamma,\eta) &&
\eea
where $\tilde \Delta$ is the Fourier transform of the brightness
perturbation, $\gam$ is the direction of the photon momentum,
$\mu \equiv \hat k \cdot \gam$, 
the dot indicates differentiation with respect to conformal time
and $h_{33}$ and $h$ are the 3,3 component and trace of the synchronous
gauge metric perturbation, respectively,
in the coordinate system defined by $x_3=\gam$.

On the right-hand side of equation (\ref{eqn:Boltz}) we have explicitly
separated out the dependence of the collision term on the
differential optical depth, $\dot \tau = a\sigma_T n_e({\bf x},\eta)$.
The quantity $S$ is a function of the
brightness perturbation and the electron velocity.  To first order in
the perturbation variables its Fourier transform is (neglecting the
polarization dependence of the Thomson cross-section)
\be
\label{eqn:S}
\tilde S({\bf k},\hat \gamma,\eta) =
-\tilde \Delta + \tilde \Delta_0 + 4 \hat
\gamma \cdot {\bf v}_e - {1\over 2}\tilde \Delta_2 P_2(\mu)
\ee
where $\Delta = \sum_\ell (2l+1) \Delta_\ell P_\ell (\mu)$.

If we were to allow both $\dot \tau$ and $S$ to depend on ${\bf x}$,
the Fourier modes would no longer evolve independently.  To avoid
this complication, we use an expansion in small $\dot \tau$.
Let ${\tilde \Delta}^{(0)}$ be the solution for $\dot \tau \equiv 0$ and
let ${\tilde \Delta}^{(1)} = \tilde{\Delta} -{\tilde \Delta}^{(0)}$.
Then to first order in $\dot \tau$:
\be
\dot {\tilde \Delta}^{(1)}({\bf k}) + ik\mu {\tilde \Delta}^{(1)}({\bf k})
=\int d^3x e^{i{\bf k}\cdot{\bf x}} \dot \tau({\bf x})S^{(0)}({\bf x},\hat \gamma).
\ee
which has the solution in real space:
\be
\label{eqn:Delta}
\Delta^{(1)}(\hat \gamma,{\bf x},\eta_0)= \int_{\eta_i}^{\eta_0} d\eta^\prime
\dot \tau({\bf x^\prime}) S^{(0)}({\bf x}^\prime,\hat \gamma)
\ee
where ${\bf x}^\prime= {\bf x} + \hat \gamma(\eta_0-\eta^\prime)$.
This solution can be derived with a Green function approach or simply verified
by substitution.  It can be taken to higher order, if desired, by
getting $\Delta^{(2)}$ from $S^{(1)}$, etc.  Note that for the Doppler
effect equation~\ref{eqn:Delta} is exact because $v_e$ (unlike, e.g., $\Delta$)
receives no corrections due to the optical depth.

We can now argue that the Doppler effect, due to the ${\bf v}_e$ term
in equation~\ref{eqn:Delta}, is the most important inhomogeneous effect.
The dominance of this effect is due to the fact that by the time of
reionization, velocities have grown substantially as they react to the
gravitational field.  On the relevant length scales, the rms peculiar
velocities are about $10^{-3}$.  In contrast, the damping of the
anisotropy is due to the $-\Delta$ term in $S$, which has not grown
from its primordial value of $10^{-5}$.  Similarly, polarization is
sourced by the quadrupole moment which is also still at the $10^{-5}$
level at the epoch of reionization. Therefore we expect these
contributions to be down by two orders of magnitude in amplitude, or
four in power.

If we break up the correlation function into components according
to the expansion in optical depth:
\be
C(\theta) = C^{(0)}(\theta)+2C^{(01)}(\theta)+C^{(1)}(\theta)
\ee
where $C^{(0)}$, $C^{(1)}$ and $C^{(01)}$ are the correlations
between the two zeroth order terms,
between the two first order corrections
and between the first order and zeroth order terms, respectively.

For the Doppler effect, the (exact) correction to the homogeneous
case comes entirely from the $C^{(1)}$ term which
is equal to
\be
\label{eqn:CDoppler}
C^{(1)}(\theta) =
\int_{\eta_i}^{\eta_0}d\eta_1
\int_{\eta_i}^{\eta_0} d\eta_2 \langle \left(\dot \tau \hat \gamma_1 \cdot
{\bf v_e}\right)_1 \left(\dot \tau \hat \gamma_2 \cdot
{\bf v_e}\right)_2  \rangle
\ee
where the numeric subscript on the quantities in parentheses means evaluate
at, e.g., $({\bf x}+\hat \gamma_1(\eta_0-\eta_1),\eta_1)$.

 From equation ~\ref{eqn:CDoppler} we see that 
the relevant quantity for the CMB 2-point autocorrelation function,
is the two-point function of the product,
$ \dot \tau \hat \gamma \cdot {\bf v}_e$, integrated over the lines
of sight.  As pointed out by Knox et al. (1998) it is this
latter 2-point function which must be
calculated from either simulations or analytical models
of the reionization process.

\subsection{Results from Simple Models}

A very simple toy model was used by Gruzinov and Hu (1998, hereafter GH) and
by Knox et al. (1998, hereafter KSD) for investigating the effect of IHR on
the power spectrum.  In this model, independent sources turn
on randomly and instantaneously ionize a sphere with comoving
radius $R$, which then remains ionized.  The rate of source creation
is such that the mean ionization fraction is zero prior to
$z_{\rm ion}+\delta z$ and then rises linearly with redshift to unity
at $z_{\rm ion}$.  GH derived the following approximate
expression for the resulting power spectrum:
\bea
\label{eqn:GH}
{l^2 C_l \over 2\pi} &=& A l^2 \theta_0^2 e^{-l^2 \theta_0^2/2}, \nonumber \\
\theta_0 & \equiv & {R \over \eta_0 - \eta_{\rm ion}}, \nonumber \\
A &=& {\sqrt {2\pi} \over 36} \tau_0^2 <v^2> R/\eta_0 \delta z
\left(1+z_{\rm ion}\right)^{3/2}
\eea
where $\langle v^2\rangle $ is the variance of peculiar velocities
and $\tau_0 \equiv \sigma_T \eta_0 n_0$ and $n_0$ is the number
density of free electrons today.  The qualitative shape of this
power spectrum is simple to understand.  On large scales it is
the shape of a white noise power spectrum ($C_l = $ constant),
due to the lack of correlations between the patches.  At small
scales there is an exponential cutoff corresponding to the 
angular extent of the patches themselves, since they have
no internal structure.

The exact calculations for this model by KSD result in a very similar
shape but with the power reduced by a factor of two and the location
of the peak shifted to larger $l$ by 30\%.  Thus we can write down
very simple, and accurate, equations for the location of the maximum,
which are those of GH with the prefactors modified to fit the results of
KSD:
\bea
\label{eqn:dopplermax}
\left({l^2 C_l \over 2\pi}\right)_{\rm max} &=& {\sqrt {2\pi} \over 36e}
\tau_0^2 <v^2> R/\eta_0 \delta z
\left(1+z_{\rm ion}\right)^{3/2}, ~{\rm and} \nonumber \\
l_{\rm max} &= &{1.8 \eta_0 \over R} \left[1-(1+z_{\rm ion})^{-1/2}\right].
\eea

Note that for sufficiently large $R$, velocities are not coherent within
a patch and equations~\ref{eqn:GH} and~\ref{eqn:dopplermax} break
down; cancellations
once again become important.  However, this
only happens for $R \ga 10$ Mpc since the velocity correlation
function only becomes negative for separations greater than
about $30 h^{-1}$ Mpc (KSD).  As long
a $R$ is less than 10's of Mpc, the power scales linearly with $R$.

\begin{figure}[bthp]
\plotone{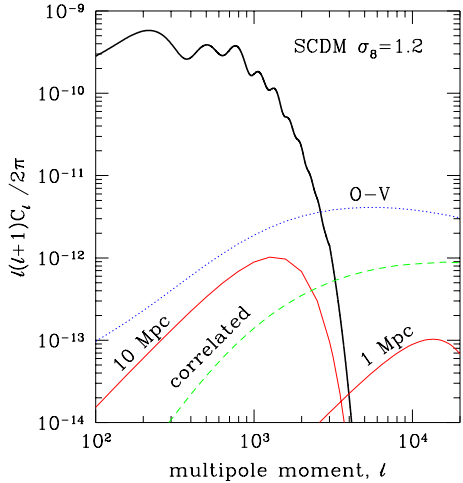}
\caption[]{The thick solid line is the angular power spectrum of the
primary CMB anisotropy in the standard cold dark matter (SCDM) model,
normalized so that the fractional rms fluctuations of the 
mass in $8 h^{-1}$ Mpc spheres is $\sigma_8=1.2$.  The other
curves are contributions to the power spectrum expected from different 
models of IHR.  The light solid curves are for the uncorrelated model 
(see equation~\ref{eqn:GH}) with $z_{\rm ion} = 5.6$ and $\delta z = 5$
with comoving patch radii of $R= 10$ Mpc and $R = 1$ Mpc.  The
dashed curve is for the correlated model of KSD which achieves
full reionization at $z_{\rm ion} = 26$.  The underlying model
is spatially flat CDM with $\Omega_b = 0.1$ and $h=0.5$.
The expected O-V effect for this same model with the same reionization
redshift is also shown, calculated using the formalism of 
Jaffe \& Kamionkowski (1998). 
}
\label{fig:pspec} \end{figure}

Aghanim et al. (1996, hereafter ``A96'') used a slightly more
complicated model in which there was a distribution of patch sizes due
to an assumed distribution of mini--quasar luminosities.  The larger
patches dominate at low $l$ and the cutoff at higher $l$ is much
softer than in the single patch case, due to the existence of the
smaller patches.  For an order-of-magnitude calculation, their largest
patches have $R \simeq 10$~Mpc and reionization proceeds from $z_{\rm
  ion}+\delta z = 10.6$ to $z_{\rm ion}=5.6$.  For this simplification
of the A96 model we find the curve in
figure~\ref{fig:pspec} which has $\left({l^2 C_l \over
    2\pi}\right)_{\rm max} \simeq 1 \times 10^{-12}$ and $l_{\rm max}
\simeq 1300$.

This amplitude is just large enough to possibly be a source of
systematic error in the determination of cosmological parameters from
{\it MAP} data, which has sensitivity out to $l \simeq 1000$.  To
be more precise, {\it MAP} can determine the power in a broad band
centered on $l = 800$ with width $\delta l = 400$ to about 0.5 \%.
Thus, given the primary signal assumed in the figure, we may expect
signals on the order of $5 \times 10^{-13}$ to be significant
contaminants.  This IHR signal from a 10 Mpc patch would be an even
more noticeable effect in the {\it Planck} data since {\it Planck} has
sensitivity out to about $l \simeq 2000$ and can determine the power
in a broad band centered on $l = 1500$ with width $\delta l = 1000$ to
about 0.2 \%.

The quasar luminosity function (QLF) is relatively well measured in
the optical below redshift $z<5$, with a typical average optical
luminosity of $10^{46}~{\rm erg~s^{-1}}$. In order to estimate the
expected typical HII patch size around the observed quasars, one needs
to know their lifetime (or "duty--cycle").  Taking the lifetime of
quasars to be universal, and near the Eddington time
($\sim5\times10^7$ yr), and further ignoring recombinations, one
obtains a comoving radius of $\sim$10--20Mpc.  As seen above, this
would be large enough for its effects on the CMB to be detectable
by {\it Planck} and maybe {\it MAP}.

We emphasize however, that the typical quasar luminosity at $z=10$ can
be much lower than at $z\sim 3$.  In hierarchical structure formation
models, the nonlinear mass--scale is $\sim3$ orders of magnitude
smaller at $z=10$ than at $z\sim3$. With a linear scaling between
quasar luminosity and halo mass, this would reduce the average
patch--size by a factor of $\sim10$.  In addition, the duty--cycles of
quasars are not well established.  In theoretical models where the QLF
is derived by associating quasars with collapsed dark halos (in the
Press--Schechter formalism), a typical lifetime of $\sim5\times10^7$
yr would generically lead to an overestimate of the number of observed
quasars (e.g. Small\& Blandford 1992).  To avoid these
overpredictions, one needs to assume that either (1) the quasar shines
at a small fraction of the Eddington luminosity, or (2) the efficiency
of black hole formation, expressed as $M_{\rm bh}/M_{\rm halo}$, is 2
orders of magnitude lower than suggested by nearby galaxies (Magorrian
et al.  1998), or (3) the black hole masses grow in an optically
inactive phase (e.g. Haehnelt, Natarajan \& Rees 1998).
Alternatively, one can fit the observed QLF by assuming that the
quasar lifetime is $\sim$100 times shorter than the Eddington time
(Haiman \& Loeb 1997).  In the latter case, the HII patch sizes would
be reduced to $R \lsim1$ Mpc (i.e. reionization would correspond to a
larger number of smaller patches).

According to equation~\ref{eqn:dopplermax}, patches smaller than a few
Mpc result in power spectra with smaller maxima at values of $l$ that
are too high, and well outside the range of sensitivity of {\it MAP}
and {\it Planck}.  Even for patches as large as 1 Mpc, one can see
that, at least for reionization at the extreme low end of the redshift
range, their effect is negligible (see figure~\ref{fig:pspec}).  A
detection of an IHR signal in the CMB with a peak could have important
implications for quasar evolution theories, e.g. by ruling out models
that lead to short duty cycles and small patch sizes.  A detection
might also prove that reionization was caused by mini--quasars, rather
than mini--galaxies, since for the mini-galaxy models, the patch sizes
are expected to be only a few hundred kpc.  However, in light of the
small expected patch sizes, it seems unlikely that {\it MAP} or {\it
  Planck} reaches the required angular resolution for such a
detection.

The above calculations of the IHR power spectrum 
all assumed that the ionized regions are uncorrelated.
However, the pattern of density fluctuations will affect the pattern
of ionized regions and therefore correlations in the density field
should give rise to correlations in the ionization fraction field.
Determining what these correlations should be is complicated by the
fact that the density field affects the ionization field in two
different ways: by the fact that the sources are most likely located
in high density regions and the fact that the recombination rates are
highest in the densest regions.

To take the former effect into account, KDS gave the ionization field
the same correlation structure as that of the ``high-peaks'' of the
density field.  These high-peaks are regions where the fractional
density perturbation, $\delta = {\delta \rho \over \rho}$, smoothed on
the appropriate scale, exceeds the critical value for spherical
collapse.  The smoothing scale was taken to be that corresponding to
$10^{8}{\rm M_\odot}[(1+z)/11]^{-3/2}$, since objects below this size
will not be able to cool sufficiently to fragment and form stars
(Haiman, Rees \& Loeb 1997).  One further assumption was that the
sources ionized a region $E$ times larger than that from which they
collapsed where the efficiency factor is time-dependent and given by
(Haiman \& Loeb 1997, Haiman \& Loeb 1998a) $E=7 \times 10^5
(\eta/\eta_0)^6$.  For this rather high choice for the efficiency, 
reionization is complete at $z_{\rm ion}=26$.

The resulting spectrum is shown as the dashed line in 
figure~\ref{fig:pspec}.
The correlations between the patches---which are individually
only hundreds of kiloparsecs in radius---drastically 
alter the low $l$ shape. 
Even given the high redshift of reionization in this scenario, the
corresponding uncorrelated model of equation~\ref{eqn:GH} would 
predict orders of magnitude less power at $l \sim 10^3$ to $10^4$.  

If this signal were in the data, 
but not in ones model of the data, it would lead
to a bias in the determination of cosmological parameters.  
KSD estimated the bias that would occur in a  
nine parameter fit, assuming the primary signal was that of
standard cold dark matter.
For Planck, the systematic in the estimate of the baryon density
was equal to the statistical error.  For all the other
parameters the systematic error was less than the statistical error.

\subsection{Discussion}

If the mini--quasar scenario is correct, with bright, long-lived
quasars, then the Doppler effect will lead to a significant
contamination of the {\it Planck} data and possibly a marginally
important one for {\it MAP} data.  For the more likely mini--galaxy
scenario, if one ignores the correlations in the ionized patches, the
contamination is completely irrelevant for both {\it MAP} and {\it
  Planck}.  However, correlations may indeed be important, and in one
attempt to take them into account, we have seen that for an extreme
mini--galaxy scenario, the contamination is marginally significant
for {\it Planck}.

Work by Oh (1999) supports the idea that the ionized patches
should be correlated.  Oh has modeled the ionizing sources
as characterized by a mean separation (given 
by Press-Schechter theory), an isotropic attenuation length which
specifies how far the ionizing radiation propagates, and a power-law
correlation function (appropriate for these highly-biased objects).
Placing sources in a box and working
out the resulting flux densities he can then calculate the correlation
function of the flux density.  It is indeed correlated, with significant
(0.2) correlation at a comoving separations of 10 Mpc.  Although
a clumping factor for the IGM from Gnedin \& Ostriker (1997) 
is used to estimate the attenuation length, all propagation from
the sources is taken to be isotropic, and hence this work does not
yet include the influence of shielding and shadowing.

As one can see in Fig.~\ref{fig:pspec}, the contribution from IHR may be
sub-dominant to that from the Ostriker-Vishniac (O-V) effect, as it is for
all three models of IHR shown.  The O-V contribution
here is calculated using the formalism of Jaffe \& Kamionkowski (1998,
hereafter ``JK98'').  One must keep in mind that as a second-order
effect, it is highly sensitive to the normalization.  The curve here
is COBE-normalized ($\sigma_8 = 1.2$); a cluster normalization of
$\sigma_8 = 0.6$ would reduce the $C_l$ by a factor of 16.  It is also
calculated for the very early reionization redshift (and resulting
high optical depth) of $z_{\rm ion} =
26$.  However, the amplitude is not very sensitive to $\tau$ because
much of the effect actually comes from more recent times---the
dropping number density of electrons must compete with the growth in
the density contrasts and peculiar velocities(Hu \& White 1995, JK98).  
Of course, the separation into
IHR and O-V is artificial, and efforts to understand
the contribution from reionization are surely going to benefit
from treating them simultaneously.  Indeed, 
the 2-point function of $\dot \tau \hat \gamma \cdot {\bf v}_e$,
measured from a simulation, would include both effects.

Gravity is the only other late-time,
frequency-independent, influence on the CMB photons, causing
lensing and the Rees-Sciama effect 
(Rees \& Sciama 1968).  The
angular-power spectrum from the Rees-Sciama effect has
been calculated by Seljak (1995) and is substantially sub-dominant to
either the primary spectrum or the O-V effect
at all angular scales.  Gravitational lensing results 
in a smearing of the primary CMB angular power spectrum and has also
been calculated by Seljak (1996).  Uncertainties in this 
calculation may indeed be important for any attempt to recover
the contribution from the IHR and Ostriker-Vishniac effects.

\section{Conclusions}

The principal unanswered questions about reionization are: what type
of sources caused it, and at what redshift did it occur?  These
questions have been addressed both theoretically and observationally.
 From a theoretical point of view, the two leading candidate sources
are an early generation of stars (``mini--galaxies''), or massive
black holes in small halos (``mini--quasars'').  It is possible to
estimate the efficiencies with which objects form in the earliest
collapsed halos: from the metals in the Ly$\alpha$ forest, and from
the evolution of the quasar luminosity function at $z<5$,
respectively.  However, the uncertainties in these efficiencies are
still too large to allow definite predictions.  The expected
reionization redshift, depending on the type of source, cosmology,
power spectrum, and a combination of the efficiency factors, is
between $7\lsim z \lsim 20$ in homogeneous models.  A further
complication is the inhomogeneous nature of reionization: the sources
of ionizing radiation are likely clustered, and embedded in complex
dense regions, such as the filaments and sheets seen in 3D
simulations. At the same time, the gas to be ionized likely has
significant density fluctuations.

Further theoretical progress on the problem will likely come from 3D
simulations.  Such simulations must be able to follow the propagation
of a non-spherical ionization front into a medium that has large
density fluctuations, as well as opaque clumps of absorbing material.
In a homogeneous medium populated by randomly distributed sources, the
background radiation would have negligible fluctuations (e.g. Zuo
1992).  However, because the density inhomogeneities absorb the UV
flux of the ionizing sources, and re-emit diffuse ionizing photons as
they recombine, the ionizing background will likely become
significantly inhomogeneous and anisotropic (Norman et al. 1998).  The
first numerical algorithms to deal in full detail with these problems
have been proposed by Abel, Norman \& Madau (1998) and Razoumov \&
Scott (1998) (see also Gnedin 1998).

The reionization redshift can also be constrained observationally,
with a resulting uncertainty that is comparable to the theoretical
range: $6\lsim z \lsim 40$.  On both ends of this redshift range,
observational progress is likely in the near feature. If new
Ly$\alpha$ emitters are discovered at higher redshifts, this would
improve the lower bound.  Conversely, as the CMB data are collected,
the constraint on the electron scattering optical depth will improve,
tightening the upper bound.

We have seen how the optical depth to Thomson-scattering, and hence
the epoch of reionization, can be constrained by determining the
amount of damping in the CMB temperature anisotropy and by detection
of the polarization contribution at very large angular scales.  The
spatial inhomogeneity of the reionization process gives us a further
opportunity to probe in more detail the epoch of reionization.  For
this probe to become reality, further development of theoretical
predictions is required, as well as more sensitive measurements at arc
minute and smaller angular scales.

Fortunately, the inhomogeneity of reionization is unlikely to spoil
our ability to interpret the primary CMB anisotropy, to be measured
with exquisite precision over the next decade, although this
may not be the case if reionization did indeed occur via mini--quasars.  
For the mini--galaxy models, even when reionization occurs at the
very high end of the allowed redshift range, very little
signal is produced at the relevant angular scales.  
Further progress in these predictions is likely to be
motivated by the desire to understand the spectrum at $l \ga 3000$ as a
means of probing the end of the dark ages, rather than at $l \la
3000$, as a possible contaminant of the primary signal.

\acknowledgments

We are grateful to our collaborators S. Dodelson, A. Loeb, M. Rees,
and R. Scoccimarro for contributing to our understanding of this subject,
to A. Jaffe for supplying the ``O-V'' effect curve in
Fig.~\ref{fig:pspec}, and to M. Norman 
for supplying Fig. 2.  ZH was supported at Fermilab by the DOE and the 
NASA grant NAG 5-7092.

{\bf References}

\noindent Abel, T., Norman, M. L., \& Madau, P. 1998, ApJL, submitted,
           preprint astro-ph/9812151

\noindent Aghanim, N., D\'esert, F. X., Puget, J. L., \& Gispert, R.
           1996, A\&A, 311, 1

\noindent Arons, J., \& Wingert, D. W. 1972, ApJ, 177, 1

\noindent Baltz, E. A., Gnedin, N. Y., \& Silk, J. 1998, ApJL, 493, 1

\noindent Barkana, R., \& Loeb, A. 1999, ApJ, submitted, preprint astro-ph/9901114

\noindent Baugh, C. M., Cole, S., Frenk, C. S. \& Lacey, C. G.
           1998, ApJ, 498, 504

\noindent Bouchet, F. R., Prunet, S., Sethi, K. S., preprint astro-ph/9809353

\noindent Carr, B. J., Bond, J. R. \& Arnett, W. D. 1986, ApJ., 306, L51

\noindent Conti, A., Kennefick, J. D., Martini, P., \& Osmer, P. S. 1999,
           AJ, in press, astro-ph/9808020

\noindent Couchman, H. M. P. \& Rees, M. J. 1986, MNRAS, 221, 53

\noindent Dodelson, S. \& Jubas, J. 1995, ApJ, 439, 503

\noindent Elvis, M., Wilkes, B. J., McDowell, J. C., Green, R. F.,
           Bechtold, J., Willner, S. P., Oey, M. S., Polomski, E.,
           \& Cutri, R. 1994, ApJS, 95, 1

\noindent Eisenstein, D. J., Hu, W., Tegmark, M., preprint astro-ph/9807130 

\noindent Fabian, A. C. \& Barcons, X. 1992, ARA\&A, 30, 429

\noindent Fardal, M. A., Giroux, M. L., \& Shull, J. M. 1998, AJ, 115, 2206

\noindent Fixsen, D. J., Cheng, E. S., Gales, J. M., Mather, J. C.,
           Shafer, R. A., \& Wright, E. L. 1996, ApJ, 473, 576

\noindent Fixsen, D. J., Dwek, E., Mather, J. C., Bennett, C. L., 
           Shafer, R. A., 1998, ApJ, inpress, preprint astro-ph/9803021.

\noindent Fukugita, M. \& Kawasaki, M. 1994, MNRAS, 269,563.

\noindent Gibilisco, M. 1996, Int. J. Mod. Phys. A11, 5541,
           preprint astro-ph/9611227

\noindent Gnedin, N. Y. 1998, in Proc. of 19$^{\rm th}$ Texas Symposium
           on Relativistic Astrophysics and Cosmology, held in Paris, France,
           Dec. 14-18, 1998, Eds. J. Paul, T. Montmerle, and E. Aubourg
           (CEA Saclay), in press

\noindent Gnedin, N. Y., \& Ostriker, J. P. 1997, ApJ, 486, 581

\noindent Griffiths, L. M., Barbosa, D., Liddle, A. R. 1998, MNRAS, submitted,
           preprint astro-ph/9812125

\noindent Gruzinov, A., \& Hu, W. 1998, ApJ, in press,
           preprint astro-ph/9803188

\noindent Gunn, J. E., \& Peterson, B. A., 1965, ApJ, 142, 1633

\noindent Haehnelt, M. G., Natarajan, P., \& Rees, M. J. 
           1998, MNRAS, 300, 817

\noindent Haiman, Z., Abel, T., \& Rees, M. J. 1999, ApJ, to be submitted

\noindent Haiman, Z., \& Loeb, A. 1997, ApJ, 483, 21

\noindent Haiman, Z., \& Loeb, A. 1998a, ApJ, 503, 505

\noindent Haiman, Z., \& Loeb, A. 1998b, ApJ, in press,
           preprint astro-ph/9807070

\noindent Haiman, Z., \& Loeb, A. 1998c, invited contribution
           to the Proceedings of 9$^{\rm th}$ Annual October Astrophysics
           Conference, {\it After the Dark Ages: When Galaxies Were
           Young}, October 1998, College Park, MD, preprint astro-ph/9811395

\noindent Haiman, Z., Madau, P., \& Loeb, A. 1999, ApJ, in press, preprint
           astro-ph/9805258

\noindent Haiman, Z., \& Spaans, M. 1998, ApJ, in press,
           preprint astro-ph/9809223

\noindent Haiman, Z., Rees, M. J., \& Loeb, A. 1997, ApJ, 476, 458

\noindent Hauser, M. G. 1998, ApJ, 508, 25

\noindent Hu, E. M., Cowie, L. L., \& McMahon, R. G.\ 1998, ApJ, 502, 99

\noindent Hu, W., \& White, M. 1995,  Astronomy and Astrophysics 315, 33

\noindent Hu, W., \& White, M. 1997, ApJ, 479, 568

\noindent Hu, W., Scott, D. \& Silk, J. 1994, Phys. Rev. D49, 648

\noindent Jaffe, A. H., \& Kamionkowski, M., Phys. Rev. D58, 1998, 
           043001

\noindent Kaiser, N. 1984, ApJ, 282, 374

\noindent Knox, L., Scoccimarro, R., Dodelson, S., 1998, Phys. Rev. Lett.,
           81, 2004

\noindent Knox, L., 1998, preprint astro-ph/9811358

\noindent Loeb, A., \& Haiman, Z. 1997, ApJ, 490, 571

\noindent Ma, C.-P. \& Bertschinger, E. (1995), ApJ, 455, 7

\noindent Madau, P., Haardt, F., \& Rees, M. J. 1998, ApJ, submitted,
           preprint astro-ph/9809058

\noindent Madau, P., Meiksein, A., \& Rees, M. J. 1997, ApJ, 475, 429

\noindent Magorrian, J., et al. 1998, AJ, 115, 2285

\noindent Meiksin, A., \& Madau, P. 1993, ApJ, 412, 34

\noindent Miralda-Escud\'e, J. 1998, ApJ, 501, 15

\noindent Miralda-Escud\'e, J., Haehnelt, M., \& Rees, M. J. 1999,
           ApJ, submitted, preprint astro-ph/9812306

\noindent Miralda-Escud\'e, J., \& Rees, M. J. 1994, MNRAS, 266, 343

\noindent Nath, B. B., \& Bierman, P. L. 1993, MNRAS, 265, 241

\noindent Norman, M. L., Paschos, P., \& Abel, T. 1998,
           in Proc. of {\it $H_2$ in the Early Universe},  Workshop
           held in Florence, Italy, eds. E.  Corbelli, D. Galli, and
           F. Palla,  Memorie Della Societa Astronomica Italiana, p. 455

\noindent Oh, S.P., work in progress

\noindent Ostriker, J. P., \& Vishniac, E. T., 1986, ApJL, 306, 51

\noindent Ostriker, J. P., \& Gnedin, N. Y. 1996, ApJ, 472, 603

\noindent Ostriker, J. P., \& Cowie, L. L., 1981, ApJL, 243, 127

\noindent Pei, Y. C. 1995, ApJ, 438, 623

\noindent Prunet, S., Sethi, K. S., Bouchet, F. R., preprint astro-ph/9803160

\noindent Press, W. H., \& Schechter, P. L. 1974, ApJ, 181, 425

\noindent Puget, J.-L., Abergel, A., Bernard, J.-P., Boulanger, F.,
           Burton, W. B., Desert, F.-X., Hartmann, D., 1996, A\&A 308, 
           L5

\noindent Rauch, M., Miralda-Escud\'e, J., Sargent, W. L. W.,
           Barlow, T. A., Weinberg, D. H., Hernquist, L., Katz, N.,
           Cen, R., \& Ostriker, J. P. 1997, ApJ, 489, 7

\noindent Razoumov, A., \& Scott, D. 1998, MNRAS, submitted,
           preprint astro-ph/9810425

\noindent Rees. M. J., \& Sciama, D. W. 1968, Nature, 517, 611

\noindent Reimers, D., K\"ohler, S., Wisotzki, L., Groote, D.,
           Rodriguez-Pascual, P., \& Wamsteker, W. 1997, A\&A, 326, 489

\noindent Reisenegger, A., \& Miralda-Escud\'e, J. 1995, ApJ, 449, 476

\noindent Scalo, J. M. 1986, Fundamentals of Cosmic Physics, vol. 11, p. 1-278

\noindent Sciama, D. W. 1993, {\it Modern Cosmology and the Dark
           Matter Problem}, Cambridge University Press

\noindent Scott, D., \& Rees, M. J. 1990, MNRAS, 247, 510

\noindent Seljak, U. 1995, preprint astro-ph/9506048 

\noindent Seljak, U., 1996, ApJ, 463, 1

\noindent Schlegel, D. J., Finkbeiner, D. P., Davis, M., 1998, ApJ, 500, 525

\noindent Shapiro, P. R., Raga, A. C., \& Mellema, G. 1997,
           in {\it Structure and Evolution of the IGM from
           QSO Absorption Line Systems}, 13$^{\rm th}$ IAP Colloquium,
           eds. P. Petitjean and S. Charlot (Paris: Editions Frontiere),
           in press, preprint astro-ph/9710210

\noindent Shapiro, P. R., \& Giroux, M. L. 1987, ApJ, 321, L107

\noindent Shapiro, P. R., Giroux, M. L., \& Babul, A. 1994, ApJ, 427, 25

\noindent Shapiro, P. R., Giroux, M. L., \& Kang, H. 1987, in High Redshift
           and Primeval Galaxies, eds. J. Bergeron, D. Kunth, B.
           Rocca-Volmerange, and J. Tran Thanh Van (Paris: Editions
           Frontieres), pp. 501-515

\noindent Shaver, P., Windhorst, R. A., Madau, P., \& de Bruyn, A. G. 1999,
           A\&A, submitted, preprint astro-ph/9901320

\noindent Small, T. A., \& Blandford, R. D. 1992, MNRAS, 259, 725

\noindent Songaila, A., \& Cowie, L. L. 1996, AJ, 112, 335

\noindent Stebbins, A., \& Silk, J. 1986, ApJ, 300, 1

\noindent Sunyaev, R. A. 1978, in {\it Large-Scale Structure of the Universe},
           eds. M.S. Longair \& J. Einasto (Dordrecht: Reidel), p. 393

\noindent Tegmark, M., Silk, J., \& Blanchard, A. 1994, ApJ, 420, 484

\noindent Thompson, R., et al. 1998, preprint astro-ph/9810285

\noindent Tytler, D. et al. 1995, in {\it QSO Absorption Lines}, ESO
           Astrophysics Symposia, ed. G. Meylan, Springer, Heidelberg, p.289

\noindent Vishniac, E. T., ApJ, 1987, 322, 597.

\noindent Wadsley, J. W., Hogan, C. J., Anderson, S. F. 1998,
           to appear in the proceedings of "After the
           Dark Ages: When Galaxies were Young (the Universe at $2<z<5$)",
           9th Annual October Astrophysics Conference in Maryland,
           preprint astro-ph/9812239

\noindent Weinberg, D. H., Miralda-Escud\'e, J., Hernquist, L., \& Katz, N.
           1997, ApJ, 490, 564

\noindent Weymann, R.J., Stern, D., Bunker, A., Spinrad, H., Chaffee, F.H.,
           Thompson, R.I., \& Storrie-Lombardi, L.J.\ 1998, ApJL, 505, 95

\noindent Zaldarriaga, M., Spergel, D., \& Seljak, U. 1997, ApJ, 488, 1

\noindent Zhang, Y., Meiksin, A., Anninos, P., \& Norman, M. L. 1998, ApJ, 495, 63

\noindent Zuo, L. 1992, MNRAS, 258, 36

\end{document}